\documentclass{ws-procs9x6}

\def\etal {{\it et al.}}

\newcommand{\beq}[1]{\begin{equation}\label{#1}}
\newcommand{\eeq}{\end{equation}}
\newcommand{\bea}[1]{\begin{eqnarray}\label{#1}}
\newcommand{\eea}{\end{eqnarray}}

\def\ktr{\tilde{\kappa}_{\rm tr}}
\def\kop{\tilde{\kappa}_{o+}}

\def\etal{{\it et al.}}

\def\lsim{\mathrel{\rlap{\lower3pt\hbox{$\sim$}}
    \raise2pt\hbox{$<$}}}
\def\gsim{\mathrel{\rlap{\lower3pt\hbox{$\sim$}}
    \raise2pt\hbox{$>$}}}

\begin{document}

\title{KINEMATICAL LORENTZ-SYMMETRY TESTS\\ 
AT PARTICLE COLLIDERS}

\author{RALF LEHNERT}

\address{Instituto de Ciencias Nucleares,
Universidad Nacional Aut\'onoma de M\'exico\\
A.~Postal 70-543, 04510 M\'exico D.F., Mexico\\
E-mail: ralf.lehnert@nucleares.unam.mx}

\begin{abstract}
Violations of Lorentz symmetry are typically associated with 
modifications of one-particle dispersion relations. 
The physical effects of such modifications in particle collisions 
often grow with energy, 
so that ultrahigh-energy cosmic rays provide an excellent laboratory 
for measuring such effects. 
In this talk we argue that 
collisions at particle colliders, 
which involve much smaller energies, 
can nevertheless yield competitive constraints on Lorentz breaking.
\end{abstract}

\bodymatter

\section{Introduction}

One of the earliest ideas for testing Lorentz symmetry 
in the context of quantum gravity 
involves kinematical searches for modifications in one-particle dispersion relations.
Such modifications can indeed be accommodated 
in various theoretical approaches to more fundamental physics.\cite{lotsoftheory} 

For phenomenological purposes in this context,  
corrections to energy--momentum relations need to be modeled.\cite{thres}
To this end, 
it is appropriate to consider the long-distance limit 
of general Lorentz violation in underlying physics. 
This limit can be described by the Standard-Model Extension (SME).\cite{sme}
To date, 
the SME has provided the basis for numerous experimental\cite{kr}
and theoretical\cite{theory} studies of Lorentz violation.
As a consistent dynamical framework,
the SME permits the extraction of 
acceptable Lorentz-violating dispersion relations.
For example, 
the modified dispersion relation 
for $\textrm{spin-}\frac{1}{2}$ fermions in the flat-spacetime minimal SME (mSME)
is given in Ref.\ \refcite{dr};
the one for photons in the full flat-spacetime SME 
can be found in Ref.\ \refcite{fullphoton}.

These Lorentz-breaking dispersion relations
typically affect 
particle collisions, 
which can cause observable effects. 
For example, 
particle reactions 
that are kinematically forbidden in conventional physics 
can now occur above certain thresholds; 
and vice versa, 
processes that normally occur 
may be forbidden above certain energy scales 
when Lorentz violation is present.

Since the Lorentz violations 
in such dispersion relations
often grow with the particle's momentum, 
the novel observable signals
tend to be enhanced at higher energies. 
Common belief therefore holds 
that kinematical Lorentz-symmetry studies 
are best performed 
with ultrahigh-energy cosmic rays (UHECRs). 
However, 
an UHECR collision process 
involves the (modified) dispersion relations of {\em all} participating particles 
including the primary, 
but the exact type of the primary particle 
is often unknown. 
For this reason, 
it can be interesting to consider also Lorentz tests 
with particle collisions 
in a controlled laboratory environment 
at much lower energies.
In what follows, 
it is argued 
that such dispersion-relation tests of special relativity
at colliders
can yield competitive constraints 
on Lorentz breakdown.

The Lorentz tests at accelerators 
discussed below\cite{nonkin}
are particularly sensitive to the electron--photon sector of the flat-spacetime mSME. 
Currently, 
the $c^{\mu\nu}$ and $\tilde{k}^{\mu\nu}\equiv(k_F)_\alpha{}^{\mu\alpha\nu}$ 
coefficients in this sector
obey the weakest experimental constraints. 
We can therefore focus exclusively on $c^{\mu\nu}$ and~$\tilde{k}^{\mu\nu}$.

From a conceptual viewpoint, 
it is vital to note 
that $c^{\mu\nu}$ and $\tilde{k}^{\mu\nu}$ 
are physically equivalent 
in an electron--photon system. 
This equivalence stems from the fact  
that judiciously chosen coordinate rescalings 
freely transform the $\tilde{k}^{\mu\nu}$ and $c^{\mu\nu}$ parameters
into one another~\cite{km0102,collider}. 
Intuitively, 
this reflects the fact 
that one may choose to measure distances 
with a `ruler' composed of electrons ($c^{\mu\nu}=0$), 
or with a `ruler' composed of photons ($\tilde{k}^{\mu\nu}=0$), 
or any other `ruler' ($c^{\mu\nu},\,\tilde{k}^{\mu\nu}\neq0$).
We utilize this freedom 
by making the specific choice $c^{\mu\nu}=0$
(corresponding to an `electron ruler')
in intermediate calculations.
However, 
we present all final results in a scaling-independent 
(i.e., ruler-independent') way
by reinstating the $c^{\mu\nu}$ coefficient 
for generality.

The $\tilde{k}^{\mu\nu}$ coefficient possesses nine independent components:
it is traceless and symmetric. 
We will discuss its isotropic component 
denoted by $\ktr$
and the three parity-violating anisotropic components 
usually grouped into the antisymmetric $3\times3$ matrix $\kop$. 
The remaining five components describe parity-even anisotropies 
and are not discussed here.

\section{The isotropic component $\bm \ktr$}

An mSME analysis establishes that
the isotropic component of $\tilde{k}^{\mu\nu}$ 
parametrized by $\ktr$ leads to the following 
modified dispersion relation:\cite{km0102}
\begin{equation}
E_{\gamma}^2-(1-\ktr)\vec{p}\!\phantom{.}^{2}=0\;.\label{eq:dispersion}
\end{equation}
Equation~(\ref{eq:dispersion}) holds at leading order in $\ktr$,
and $p^{\mu}\equiv (E_{\gamma},\vec{p})$ denotes the photon's 4-momentum. 
Notice 
that the physical speed of light is $(1-\ktr)$,
which is different from the usual $c=1$.
In what follows, 
we treat the two cases $\ktr>0$ and $\ktr<0$ separately
because they are associated with different phenomenological signatures. 

{\em The case $\ktr>0$.}---For positive $\ktr$, 
the speed of light $(1-\ktr)$ is slower 
than the conventional value $c=1$. 
This implies in particular 
that the maximal attainable speed (MAS) of the electrons 
is greater than the speed of the photons.
In analogy to ordinary electrodynamics inside a macroscopic medium,
we expect a Cherenkov-type effect:\cite{cher}
charges moving faster than the modified speed of light $(1-\ktr)$ 
would be unstable against the emission of light.
With the modified photon dispersion relation~(\ref{eq:dispersion}), 
one can indeed show 
that electrons at energies above the threshold
\begin{equation}
E_{\rm VCR}=\frac{1-\ktr}{\sqrt{(2-\ktr)\ktr}}\,m_e=\frac{m_e}{\sqrt{2\ktr}}+\mathcal{O}\left(\sqrt{\ktr}\right)\label{vcrenergy}
\end{equation}
emit Cherenkov photons.
We remark 
that the threshold~(\ref{vcrenergy}) 
can also be obtained from 
the ordinary Cherenkov condition 
requiring that the electrons be faster 
than the speed of light $(1-\ktr)$. 

At LEP, 
where electrons attained the energy $E_{\rm LEP}=104.5\,$GeV, 
this Cherenkov effect was not observed. 
This essentially means 
that the LEP electrons must have been 
below the Cherenkov threshold $E_{\rm LEP}<E_{\rm VCR}$.
Equation~(\ref{vcrenergy}) then yields
\begin{equation}
\ktr-\frac{4}{3}\,c^{00} \leq 1.2\times 10^{-11}\;,\label{VCRbound}
\end{equation}
where $c^{00}$ has been reinstated 
for generality. 
Note that we have implicitly used the dynamical result\cite{Altschul:2008,collider}
that Cherenkov radiation must be highly efficient 
to deduce $E_{\rm LEP}<E_{\rm VCR}$ from the non-observation 
of $e\to e\,\gamma$.

{\em The case $\ktr<0$.}---The speed of light $(1-\ktr)$
is now greater than the conventional value $c=1$. 
In particular, 
all photons move faster than
the MAS of the electrons.
Paralleling the above Cherenkov case, 
one would then expect 
that the photon can now become unstable. 
An mSME calculation with the dispersion relation~(\ref{eq:dispersion})
indeed confirms that 
for photon energies above the threshold 
\begin{equation}
E_{\rm pair}=\frac{2m_e}{\sqrt{\ktr(\ktr-2)}}=\sqrt{\frac{2}{-\ktr}}m_e
+\mathcal{O}\left(\sqrt{\ktr}\right)\; ,\label{gammathres}
\end{equation}
photon decay into an electron--positron pair 
is kinematically allowed.\cite{ks08,collider}
The D0 experiment at the Tevatron 
has observed photons with energy in excess of $E_{\rm D0}=300\,$GeV, 
so $E_{\rm pair}$ must be greater than this value. 
We then arrive at the constraint
\begin{equation} 
-5.8\times10^{-12}\lsim\ktr-\frac{4}{3}\,c^{00}\;, 
\label{decaybound}
\end{equation}
where we have reinstated the electron's $c^{00}$ coefficient. 
Again, 
we have implicitly used the dynamical result\cite{ks08,collider}
that photon decay must be highly efficient 
to deduce $E_{\rm D0}<E_{\rm pair}$ from the non-observation 
of $\gamma\to e^+\,e^-$.

We finally remark 
that the one-sided limits~(\ref{VCRbound}) and~(\ref{decaybound})
have recently been improved 
by roughly three orders of magnitude 
with an alternative method involving colliders. 
The idea is that the synchrotron losses 
of charges moving on a circular path 
are highly sensitive to $\ktr$. 
Since such losses were accurately determined at LEP, 
a bound at the level of a few parts in $10^{15}$ 
has been obtained.\cite{synchrotron}

\section{The anisotropic parity-violating components $\bm \kop$ }

It can be shown 
that the $\kop$ components of $\tilde{k}^{\mu\nu}$
modify the photon dispersion relation as follows:
\beq{ModDR}
E_\gamma=(1-\vec{\kappa}\cdot\hat{p})\,|\vec{p}|+{\mathcal O}(\kappa^2)\;.
\eeq
Here, 
the three components of $\kop$ 
have been assembled into a 3-vector:
$\vec{\kappa}\equiv((\kop)^{23}, (\kop)^{31}, (\kop)^{12})$ 
and $\hat{p}\equiv\vec{p}/|\vec{p}|$.
For a given photon momentum $|\vec{p}|$, 
the photon energy $E_\gamma$ 
depends on the direction of propagation $\hat{p}$ 
exposing anisotropies;
reversing the direction of propagation 
reveals parity violation.

Consider now Compton scattering 
with the dispersion relation~(\ref{ModDR}), 
where the incoming photon and electron are counter-propagating. 
A leading-order mSME calculation then establishes 
that the Compton edge (CE), 
which is the maximal energy of the backscattered photon, 
is\cite{graal}
\beq{modCE}
\lambda'\simeq\lambda_{\rm CE}
\left[\,
1 + \frac{2\,\gamma^2}{(1 + 4\, \gamma\, \lambda\, /\, m)^2}\,\vec{\kappa}\cdot\hat{p}
\,\right]\,.
\eeq
Here, 
$\lambda_{\rm CE}$
denotes the conventional CE energy, 
$\gamma$ is the relativistic boost factor of the incoming electron, 
and $\lambda$ the magnitude of the incoming photon 3-momentum.
It follows 
that in present context, 
the CE depends on the direction $\hat{p}$ of in the incoming electron 
(i.e., $-\hat{p}$ for the incoming photon). 
In a terrestrial particle collider, 
the direction $\hat{p}$ 
changes constantly due to the rotation of the Earth. 
According to Eq.~(\ref{modCE}), 
this should lead to sidereal variations in the CE. 
Such variations have not been observed 
at ESRF's GRAAL facility. 
This can be used to extract 
the competitive constraint\cite{graal} 
\beq{Bound}
\sqrt{[2c_{TX}-(\kop)^{YZ}]^2+[2c_{TY}-(\kop)^{ZX}]^2} 
< 1.6\times10^{-14}\,,\quad 95\,\%\;\textrm{CL}\,,
\eeq
where we have again included the electron coefficients 
for generality.
Similar limits with different methods 
were recently obtained by B.~Altschul.\cite{ba10}

\section*{Acknowledgments}
The author wishes to thank Alan Kosteleck\'y
for the invitation to this stimulating meeting.
This work was funded in part 
by CONACyT under Grant No.\ 55310 
and by the 
Funda\c{c}\~ao para a Ci\^encia e a Tecnologia
under Grant No.\ CERN/FP/109351/2009.

\end{document}